\documentclass[prd,11pt,preprint,nofootinbib,showpacs]{revtex4-1}
\usepackage[mathcal]{euscript}
\usepackage[latin1]{inputenc}
\usepackage{latexsym}
\usepackage{amsmath}
\usepackage{hyperref}
\usepackage{amsmath}    
\usepackage{graphicx}   
\usepackage{verbatim}   
\usepackage{color}      
\usepackage{subfigure}  
\usepackage{hyperref}   
\usepackage{bm}
\usepackage{graphicx}
\usepackage{amssymb}
\usepackage{bbm}
\usepackage{float}
\usepackage{slashed}
\usepackage{amsfonts}
\usepackage{euscript}
\usepackage{amsmath}    
\usepackage{mathrsfs} 
\usepackage{bbding}
\usepackage{pifont}
\newcommand{\cm}[1]{}
\usepackage{cancel}

\makeindex
\begin{document}

\title{Gravitational echoes from macroscopic quantum gravity effects}

\author{Carlos Barcel\'o}
\email{carlos@iaa.es}
\affiliation{Instituto de Astrof\'{\i}sica de Andaluc\'{\i}a (IAA-CSIC), Glorieta de la Astronom\'{\i}a, 18008 Granada, Spain}
\author{Ra\'ul Carballo-Rubio}
\email{raul.carballo-rubio@uct.ac.za}
\affiliation{Department of Mathematics \& Applied Mathematics, University of Cape Town, Private Bag, Rondebosch 7701, South Africa}
\author{Luis J. Garay}
\email{luisj.garay@ucm.es}
\affiliation{Departamento de F\'{\i}sica Te\'orica II, Universidad Complutense de Madrid, 28040 Madrid, Spain}
\affiliation{Instituto de Estructura de la Materia (IEM-CSIC), Serrano 121, 28006 Madrid, Spain}

\begin{abstract}{New theoretical approaches developed in the last years predict that macroscopic quantum gravity effects in black holes should lead to modifications of the gravitational wave signals expected in the framework of classical general relativity, with these modifications being characterized by the existence of dampened repetitions of the primary signal. Here we use the fact that non-perturbative corrections to the near-horizon external geometry of black holes are necessary for these modifications to exist, in order to classify different proposals and paradigms with respect to this criterion and study in a neat and systematic way their phenomenology. Proposals that lead naturally to the existence of echoes in the late-time ringdown of gravitational wave signals from black hole mergers must share the replacement of black holes by horizonless configurations with a physical surface showing reflective properties in the relevant range of frequencies. On the other hand, proposals or paradigms that restrict quantum gravity effects on the external geometry to be perturbative, such as black hole complementarity or the closely related firewall proposal, do not display echoes. For the sake of completeness we exploit the interplay between the timescales associated with the formation of firewalls and the mechanism behind the existence of echoes in order to conclude that even unconventional distortions of the firewall concept (such as naked firewalls) do not lead to this phenomenon. 
}
\end{abstract}


\maketitle

\begin{center}
\end{center}

\tableofcontents

\section{Introduction \label{sec:intro}}

The first observations of gravitational waves originated in mergers of astrophysical black holes \cite{Abbott2016,Abbott2016b,Abbott2016c} mark the beginning of the study of strong gravitational fields by means of gravitational waves. Future observations will permit to extend present tests of general relativity to this so far unexplored regime, and in particular monitor closely the formation and evolution of astrophysical black holes. While the consensus among the community is that these observations will not present significant deviations from the patterns dictated by general relativity, it is at the same time often highlighted that the probable presence of surprises will be one of the most interesting features of this new window to the universe. On the other hand, there are (few) specific frameworks that have predicted the existence of substantial deviations from general relativistic gravitational wave patterns. 

Since almost the dawn of general relativity and quantum theory, it has been acknowledged that quantum gravity effects should affect the way general relativity works, leading to new gravitational physics (see, e.g., \cite{Einstein1916,Stachel1999}). It is however fair to say that the consensus has been for a long time that these modifications must not affect significantly the macroscopic behavior of stellar systems and black holes. Even the most dramatic consequence of putting together general relativity and quantum mechanics, the Hawking effect in black holes \cite{Hawking1974,Hawking1974b,Hawking1976}, has done nothing but to reinforce this view. In astrophysical terms, a black hole that evaporates through the emission of Hawking radiation does not differ significantly from a classical black hole over timescales that, leaving aside hypothetical micro \cite{Landsberg2001,Kanti2004,Barrau2005} or primordial \cite{Green2014} black holes, are many orders of magnitude larger than the age of the universe (given by the Hubble time \cite{Avelino2016}).

In the last two decades some people have started proposing that quantum gravity effects may affect our astrophysical view on black holes (see references in the next section). Among other things, these considerations warn of additional possibilities for the end state of gravitational collapse other than black holes, namely compact configurations that present no horizons. In the case in which the substitute of a black hole has a well-identified and at least partly reflecting surface outside (but extremely close to) its gravitational radius, one of the crucial robust smoking guns of its existence would be the presence of noticeable echoes from this surface. This stems from the remarkable fact that the echo time delay is logarithmic, leading to timescales that are mildly dependent on the placement of the reflecting surface. To our knowledge this was discussed for the first time in these terms in \cite{Barcelo2010} and later in \cite{Barcelo2014}. More recently, the existence of echoes has been discussed in numerical simulations of the scattering of wavepackets in the presence of a surface with such properties \cite{Cardoso2016,Cardoso2016b}. Quite remarkably, it has been argued that there is already some evidence of this phenomenon in the public data released by the LIGO collaboration \cite{Abedi2016}, though the validity of this claim has been swiftly questioned \cite{Ashton2016,Abedi2017}.

In any case, the collection and analysis of future observations will permit to answer unambiguously whether noticeable modifications to general relativistic gravitational wave patterns do exist. That theoretical considerations that just years ago were seen as highly speculative (in part, due to their lack of observational consequences testable with the technology available at the moment) can be now tested in experiments is noticeable. However, in order to exploit the potential of observations to discriminate between non-identical theoretical scenarios, it is important to discern their divergent peculiarities. Different theoretical scenarios have specific motivations, assumptions, properties and scope that are sharply different in some cases and should not be therefore blurred. This paper starts with a section that classifies different proposals of quantum effects beyond general relativity, with the idea of differentiating the associated observational effects. After this general discussion we focus on describing simple (but so far overlooked) robust requirements that modifications of the near-horizon classical geometry of black holes have to satisfy in order to lead to echoes in gravitational wave signals generated in the coalescence of binaries of compact astrophysical objects, with the aim to clarify certain misunderstandings. Even if simple, these requirements are tight enough to discard the existence of echoes in certain scenarios such as firewalls \cite{Almheiri2012,Almheiri2013} and better accommodate others such as black stars \cite{Barcelo2007,Visser2009} in which, moreover, they were indeed predicted \cite{Barcelo2010,Barcelo2014}.

\section{Black holes beyond general relativity \label{sec:classification}}

In classical general relativity (understood as the theory defined by the Einstein field equations sourced by a sensible matter content satisfying certain energy conditions), nothing can stop the final collapse of a stellar object once it has been triggered due to the dynamical formation of configurations dense enough. Hence at some point in the collapse the gravitational radius will be crossed; later on a singularity appears, and an entire compact and almost empty region of spacetime ends up enclosed inside an event horizon.\footnote{For simplicity we will always use the word ``singularity" in singular, although singular regions inside general classical black holes can be very complicated (and in particular, multiple); we will use our language as if we were only talking about spherically symmetric configurations.}
This scenario can be greatly affected by quantum gravitational effects. There is nevertheless no complete and accepted model in the literature up to now, but rather a set of partial approaches that sometimes share general and seemingly similar ideas, but entail very disparate physical perspectives. Our first aim in this paper is to present a classification of these approaches in a small number of categories. There is no such classification in the literature to our knowledge and, as we explain later in the paper, it will be of relevance to avoid potential confusions in the near future when new observational windows probe the regime of strong gravitational fields.

To start with, one can classify different scenarios that include quantum gravity effects beyond general relativity, depending on the placement of their onset:
\begin{itemize}
\item{Type I: this category includes all the scenarios in which quantum effects become significant before the formation of any trapping horizon (crossing of the gravitational radius).
}
\item{Type II: scenarios in this category restrain the onset of quantum effects to the moment in which the classical singularity would be about to appear, namely when high enough (Planckian) curvatures are attained.}
\end{itemize}
Now let us use this reference framework in order to explore in more detail approaches falling in these two categories.

Type I effects are typically sought for in the framework of semiclassical gravity (quantum field theory in curved spacetimes). Although some people claim that type I effects might occur \cite{Mazur2004,Mottola2010,Vachaspati2006,Vachaspati2007,Saini2015,Kawai2013,Kawai2014,Kawai2017,Ho2015,Baccetti2016}, we think that these arguments are difficult to sustain and at best might work for some specific situations but not as a matter of principle. The main argument against type I effects is well-known: one can imagine a large amount of matter placed so that it undergoes a spherically symmetric collapse; if the amount of matter is sufficiently large its density at horizon crossing (i.e., when crossing the gravitational radius) could be as small as that of air, and the curvature as small as desired. Finding a mechanism that could trigger the existence of large quantum gravity effects just before horizon crossing, eventually stopping the collapse, is the ultimate goal of these approaches (a recurrent candidate for this mechanism is the very backreaction due to the emission Hawking of radiation \cite{Greenwood2010,Kawai2013,Kawai2014,Kawai2017,Ho2015,Baccetti2016}).

In a paper by one of the present authors and other collaborators \cite{Barcelo2007}, it was shown that this argument against type I effects is flawed as stated above. However, the argument can be refined, exposing a true trouble for type I effects. Within the framework of semiclassical general relativity, it was shown that one may trigger the presence of large quantum vacuum effects just before crossing the gravitational radius if and only if the collapse proceeds much slower than expected in classical general relativity. Hence the initial classical model of gravitational collapse has to deviate substantially from free fall to permit that vacuum polarization effects kick in, instead of being just tiny perturbative corrections. As a result, it is neither just a matter of how small is the density of collapsing matter at horizon crossing, nor of how small is the curvature on this region, but the crucial ingredient is the velocity acquired by the collapsing distribution of matter just before entering its gravitational radius. If for any reason this velocity was small, non-local effects associated with quantum vacuum polarization could overcome any classical source of the field equations. So, the subtle reformulation of the above argument is: what physical mechanism could cause that something collapsing at a significant velocity and with the density of air, suddenly starts to slow down? There is no known such mechanism involving type I effects only but, as explained in more detail below and in Sec. \ref{sec:blackstars}, certain kind of type II effects open the door to a natural realization.

Type II effects are, on the other hand, widely expected to exist. These expectations come from the need to regularize the internal singularity of a classical black hole, which has been recognized before the Hawking paradigm of evaporating black holes was proposed \cite{Bardeen1968}. In this paradigm, a classical black hole would be substituted by an evaporating black hole, which is given by a perturbative correction of the classical geometry but close to the central singularity. An evaporating black hole is in astrophysical terms indistinguishable from a classical black hole except for a tiny and imperceptible evaporation. This paradigm famously led to the information loss problem, which has been 
a central engine of activity in theoretical physics for decades \cite{Hawking2005}. It is nowadays recognized that singularity regularization has much to say about the information loss problem (see, e.g., the discussion in \cite{Ori2012}). To even start solving this problem, it is important that the classical black hole geometry is replaced by a regular geometry (which may be of semiclassical or quantum nature). The most conservative candidate for this replacement are the so-called regular black holes (see, e.g., \cite{Dymnikova1992,AyonBeato2000,Bambi2013,DeLorenzo2015,Torres2016,Frolov2016} and references therein). The history of these objects can be traced back to the classic work of Bardeen \cite{Bardeen1968} (a history of this subject can be found in \cite{Ansoldi2008}; another brief history can be found in~\cite{Barcelo2015u}). In generic terms the idea is that the singular center of a black hole is replaced by a regular core. This small and regular core is by construction hidden to observers outside the trapping horizon (external observers in the following) for almost the entire evaporating lifetime, or decay time, of the black hole. These regular black holes are also in principle consistent with the Hawking paradigm: they may evaporate regularly, at least in geometrical terms \cite{Hayward2005,Frolov2014}. 

However, also for a long time the consensus has been that these modifications would not change substantially the astrophysical image of a black hole. To be precise, the external geometry defined as the patch of spacetime that is accesible to external observers (outside the trapping horizon) can be thoroughly described using classical general relativity plus tiny perturbative modifications coming from the evaporation described in semiclassical terms (except for the end of the evaporation, perhaps). In practical terms this means that there is no way to know experimentally what is really happening close to the classical singularity (nor even whether it is really there); only observers falling into the black hole crossing the horizon (internal observers in the following) would be able to access this part of spacetime (internal geometry in the following). This is, for instance, the view embodied in black hole complementarity \cite{Susskind1993}, one of the most popular and cunning proposals to solve the information loss problem ubiquitous in this view of black holes beyond general relativity. Let us mention, in connection with the following paragraph, that in our view the information loss problem of evaporating black holes has its roots in the very adherence to perturbative deviations from the classical picture of a black hole.    

In recent years, a different take on type II effects have been proposed in approaches of diverse nature, such as\footnote{See also prior discussions in \cite{Frolov1981,Hajicek2002,Ambrus2005}.} the scenario for the non-violent release of information in black holes proposed by Giddings \cite{Giddings2012,Giddings2013,Giddings2014,Giddings2016}, the Planck star scenario by Rovelli and collaborators \cite{Rovelli2014,Christodoulou2016}, or the proposal for the time-symmetric decay of black holes into white holes and subsequent formation of black stars by the present authors \cite{Barcelo2010,Barcelo2014e,Barcelo2014,Barcelo2015,Barcelo2015u,Barcelo2016}. The main claim shared by these works is that the possible non-perturbative nature of type II effects should not be disregarded, as this may lead to macroscopic (i.e., affecting the near-horizon external geometry) deviations from the classical picture of black holes. In rough terms, while type II effects are by definition triggered close to the would-be classical singularity, it has been pointed out in these works that they may be strongly non-local, being transmitted even beyond the gravitational radius. This has important consequences for the information loss problem (that depends on the particular theoretical scenario considered). A crucial difference though between these different proposals is the timescale in which these effects are able to modify the external geometry. This time scale has a determinant impact on the astrophysical behaviour of the resulting object.

These considerations lead to a further division within type II effects:
\begin{itemize}
\item{Type IIA: the new features of black holes beyond general relativity may be of importance for the global picture of black holes but are not relevant for observations of astrophysical black holes. Either these effects modify only the internal geometry, or potential modifications of the external geometry of black holes are completely negligible for the timescales that are available observationally. The semiclassical picture of evaporating black holes is an excellent approximation in all these scenarios.
}
\item{Type IIB: non-perturbative effects can affect the external geometry of black holes and appear in short timescales that fall within observational timescales, having an important impact on the nature of astrophysical black holes. The semiclassical picture of evaporating black holes does not capture all the relevant physics. In stark contrast with type I and type IIA effects, this implies the falsifiability of these effects in future observations that resolve these timescales.}
\end{itemize}
For clarity, under these definitions, the black-hole complementarity paradigm, Bardeen's regular black holes and Planck stars\footnote{Let us stress that this classification is made for stellar mass black holes. Planck stars lead to non-perturbative modifications that can be phenomenologically relevant only for primordial black holes of low enough mass \cite{Barrau2014}.} enter into the type IIA category.  To our knowledge, the only members in the type IIB category are Gidding's scenario for the non-violent transfer of information, and the proposal by the present authors that is explained in more detail in the paragraph below and Sec. \ref{sec:blackstars}.

One of the most important consequences of type IIB effects is that black holes as described in general relativity are not the necessary end state of gravitational collapse. Indeed, existing analyses support that a completely different end state may be reached. In the scenario discussed previously by us, the initial trapping horizon produced when the stellar structure collapses lasts only for an extremely short time interval as seen by asymptotic observers (fractions of milliseconds for figures taken from neutron stars), being disrupted by a time-reversal bounce of the stellar structure back to its initial radius (roughly the radius of a proto-neutron star). This leads to a series of rapid bounces after which, due to dissipation, the system should settle in an object very different from an evaporating black hole, namely an object filled with matter and with no horizons whatsoever. The particular mechanism proposed to lead to the stabilization of this structure makes it correspond to a black star as defined in \cite{Barcelo2007,Visser2009} (though the formation mechanism involves type IIB instead of type I effects). We recognized that this transient will lead to a characteristic gravitational wave signature whenever the initial conditions depart from pure sphericity. It will present a periodic and dampened structure, similar to a series of reverberations, with a typical timescale given by the bouncing time (which is proportional to the Schwarzschild mass) \cite{Barcelo2014,Barcelo2015u}.

From now on we will use the term ``Black and Ultra-Compact Horizonless Objects" (BUCHOs), of which black stars are an example, to refer to substitutes of astrophysical black holes that do not have horizons but are on the verge of it. It is important to stress that researchers defending type I effects are in most cases directly proposing the existence of BUCHOs.\footnote{Note that exotic stars such as boson stars \cite{Liebling2012}, quark stars \cite{Ivanenko1965,Itoh1970}, and the like, do not qualify as BUCHOs in the sense that their compacities are generally much smaller (for instance, the minimum radius for stable boson stars is around $3M$ \cite{Cardoso2016b}) and, in particular, there is no need to invoke quantum gravity effecs to make these structures stable.} This does not mean that the physics associated with the objects themselves, once formed, would necessarily be the same. In any case, they will be clearly differentiated when monitoring the physics of the collapse process itself or, in other words, the transients before the stabilization in the BUCHO structure. That is why it is important to maintain this differentiation.\footnote{There are still other researchers that argue for black hole substitutes but taking as motivation some ``disturbing'' properties of classical black holes (e.g., the complete stop of time with respect to the asymptotic region) \cite{Chapline2000}. But these characteristics are absent in any regular model of evaporating black hole (in the Bardeen sense explained above in this section). Hence we do not think that these characteristics are compelling enough to be considered by themselves a strong motivation to look for substitutes of evaporating black holes.} This classification highlights that the existence of BUCHOs itself cannot be understood naively as pointing to where quantum gravity effects happens to occur, but that this degeneracy can be broken by a detailed experimental analysis.  

The differences between these categories can be sharply summarized as follows:
\begin{itemize}
\item{Type I: quantum effects prevent not only the formation of singularities but also of trapping horizons of any sort, stopping abruptly the collapse in regions of spacetime with arbitrarily low curvature and leading directly to the formation of BUCHOs.}
\item{Type IIA: singularities are cured but trapping horizons are formed. Quantum gravity effects preserve the external geometry and, in particular, the existence of trapping horizons for extremely long timescales (many orders of magnitude greater than the Hubble time) for astrophysical black holes, making them essentially stationary during these huge periods of time. In practice, there are no experiments realized outside the horizon that are sensitive to potential deviations from the perturbative description of the external geometry that embodies the paradigm of semiclassical evaporation.}
\item{Type IIB: singularities are cured and trapping horizons are formed. The external (near-horizon) geometry is significantly modified in short timescales by non-perturbative quantum gravity effects. Experiments realized outside the horizon are sensitive to these macroscopic deviations from the external black hole geometry. In certain approaches, horizons are transitory and relaxation leads to the formation of BUCHOs (black stars) instead of black holes.}
\end{itemize}
%

\section{Gravitational echoes in compact binary coalescences \label{sec:scales}}

In this section we focus on the properties of the coalescence of binaries of black holes including quantum effects beyond general relativity, as a particular example in which to show the usefulness of the previous classification in order to extract the phenomenological implications of different approaches, avoiding potential confusions between them.

In the purely classical picture, the end state of the coalescence is assumed to be a Kerr black hole with certain mass and spin parameters $M$ and $a=J/M$. For simplicity, in the expressions below only the zeroth order in the dimensionless spin parameter $\chi=a/M\in[0,1]$ will be explicitly written. Higher orders can be explicitly evaluated (leading just to irrelevant changes in dimensionless numbers). All timescales below are measured in terms of the Killing time $t$ or, in physical terms, by the clocks of asymptotic observers at $r\rightarrow\infty$ \cite{Visser2007}. Natural units $c=G=\hbar=1$ are used.

The possible existence of echoes in the late-time gravitational wave signal originated in mergers of black holes beyond general relativity is the result of the convolution of two ingredients. Let us stress that this phenomenon is independent from the unavoidable existence of reverberations in transients during the very formation of BUCHOs in certain type IIB scenarios \cite{Barcelo2014,Barcelo2015u}, which is explained in Secs. \ref{sec:classification} and \ref{sec:blackstars}. This section deals only with the physics associated with the merger event of two compact objects, but it is important to keep in mind that the formation of these individual objects (or, once formed, their properties as single entities \cite{Akhmedov2016,Brustein2017}) is a potential source as well of an equally interesting phenomenology.

The first ingredient is the existence of an angular momentum barrier in which part of the gravitational waves generated in the merger event are reflected back towards the newly formed black hole. The value of the radius $r_0$ that marks the peak of the angular momentum barrier for different quasi-normal modes with indices $(l,m)$ can be obtained solving a sixth-order polynomial equation \cite{Yang2013} given by
\begin{equation}
2r_0^4(r_0-3M)^2+\mathscr{O}(\chi^2)=0.
\end{equation}
Hence for a Schwarzschild black hole ($\chi=0$), $r_0=3M$.

The existence of an angular momentum barrier depends only on the form of the Kerr geometry at a macroscopic distance from the gravitational radius (typically $r_0-2M\simeq M$). In order to obtain echoes, however, an additional element is needed. This is the existence of a surface outside the horizon that reflects waves with the frequencies produced in the primary emission of gravitational waves. Without the existence of such a surface, or if it lacks reflective properties (or if it is located inside the black hole), gravitational waves reflected back to the black hole will be absorbed. This reflective surface is placed in a radius $r_{\rm s}=2M+\Delta$ in which its surface is greater than the surface of the black hole by an amount proportional to the Planck area (as explained below, the particular value of the dimensionless proportionality factor is irrelevant for this phenomenon to exist).

These two ingredients lead to two different timescales. The first timescale $\mathscr{T}_{\rm s}$ is the typical time in which the surface and/or its reflective properties are developed. In principle it could be $\mathscr{T}_{\rm s}>0$ if there is some formation or relaxation time associated with the creation of this surface (as it would happen for certain distortions of the concept of firewalls, as explained in Sec. \ref{sec:firewalls}). Of course, $\mathscr{T}_{\rm s}=\infty$ for classical black holes. On the other hand, if a reflective surface exists continuously before and after the merger of black holes (i.e., if astrophysical black holes were really BUCHOs with a reflective surface), it would be $\mathscr{T}_{\rm s}=0$.

The second timescale is given by the time that it takes for the waves scattered inwards to be reflected on the reflective surface (if already developed) and travel back to the angular momentum barrier at $r_0$. This ``echo'' timescale $\mathscr{T}_{\rm e}$ is a completely geometric notion and can be evaluated straightforwardly using the Kerr metric to be
\begin{equation}
\mathscr{T}_{\rm e}=2\left[r_0-(2M+\Delta)+2M\ln\left(\frac{r_0-2M}{\Delta}\right)\right]+\mathscr{O}(\chi^2)\simeq 4kM\ln(M)+\mathscr{O}(\chi^2).\label{eq:timescale}
\end{equation}
The relevant piece in the equation above is the zeroth order in the dimensionless spin parameter $\chi$. This piece gives the echo timescale in the Schwarzschild geometry \cite{Barcelo2010,Barcelo2014}, which is given by twice the proper distance between $r_0$ and $r_{\rm s}=2M+\Delta$ in this geometry:
\begin{equation}
2\int_{r_{\rm s}}^{r_0}\frac{\text{d}r}{1-2M/r}=2\left[r_0-(2M+\Delta)+2M\ln\left(\frac{r_0-2M}{\Delta}\right)\right].
\end{equation}
For $\Delta\ll 2M$ small enough the logarithm provides the leading contribution, which is the one displayed in the second term on the right-hand side of Eq. \eqref{eq:timescale}. There, $k$ is a $\mathscr{O}(1)$ dimensionless constant, the value of which depends on the particular position of the reflective surface outside the horizon (for instance, $k=1$ for $\Delta=1$, or $k=2$ for $\Delta=1/M$). It is the logarithmic behavior of Eq. \eqref{eq:timescale} that makes large modifications of the position of the reflective surface correspond to $\mathscr{O}(1)$ modifications of $\mathscr{T}_{\rm e}$. On the other hand, corrections from finite values $\chi\in[0,1]$ might be important for experimental studies (if enough precision is attained), but are irrelevant for the main features of this picture.

A necessary condition for echoes to exist (see Fig. \ref{fig:fig1} below) is that
\begin{equation}
\mathscr{T}_{\rm s}<\frac{\mathscr{T}_{\rm e}}{2}.\label{eq:maincond}
\end{equation}
This condition guarantees that the reflective surface is completely developed once the backscattered waves reach the proximities of the gravitational radius.

\vspace{0.5cm}

\begin{figure}[h]%
\vbox{ \hfil  \includegraphics[width=0.8\textwidth]{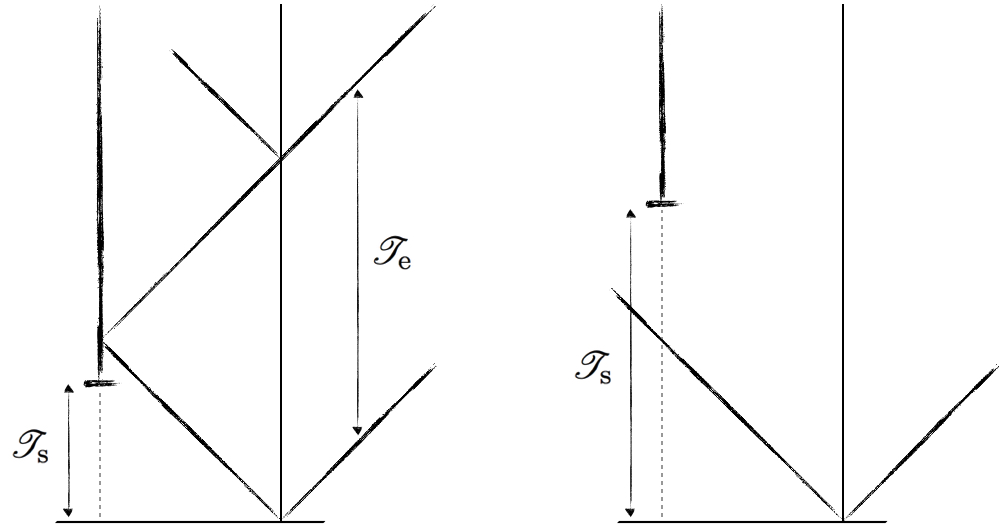}\hfil}
\bigskip%
\caption{Diagram showing the interplay between the two timescales involved: the timescale $\mathscr{T}_{\rm s}$ of formation of the reflective surface and the echo timescale $\mathscr{T}_{\rm e}$.}
\label{fig:fig1}%
\end{figure}%

We will highlight later the constraints that Eq. \eqref{eq:maincond} imposes on theoretical models that aim to account for the existence of a reflective surface leading to echoes. Models with a formation time in which the reflective surface develops dynamically will (or will not) lead to echoes depending on whether Eq. \eqref{eq:maincond} is satisfied. Quite trivially, Eq. \eqref{eq:maincond} is not satisfied for classical black holes (for which $\mathscr{T}_{\rm s}=\infty$), as it should be.

First of all, Eq. \eqref{eq:maincond} has a broader implication with deep conceptual meaning. It implies that echoes can exist only if modifications of the near-horizon external geometry of black holes arise in extremely short timescales (typically milliseconds for stellar mass black holes). These modifications are therefore perceivable by observers that remain outside the black hole horizon. 
This would imply the existence of non-perturbative deviations from the laws of classical general relativity at macroscopic distances and therefore affecting the external geometry of black holes, in regions of spacetime in which curvature invariants are small, and in short timescales [shorter than Eq. \eqref{eq:timescale}]. This observation fits nicely with our previous classification of quantum gravity effects beyond general relativity, ruling out the existence of echoes in any approach in the type IIA category. 

Before entering in more details let us focus our attention on the cases in which Eq. \eqref{eq:maincond} is trivially satisfied, given by $\mathscr{T}_{\rm s}=0$ (the complementary of $\mathscr{T}_{\rm s}=\infty$ associated with a classical black hole). These cases represent the most natural scenario for the existence of echoes and, in physical terms, correspond to BUCHOs with a reflective surface instead of black holes. As explained in more detail in the next section, it is not mandatory that all the approaches within type I and type IIB categories may lead to this structure (let us stress the need for reflective properties in the relevant range of frequencies) but, in any case, only approaches in these two categories have the potential to do so.

\subsection{Black and ultra-compact horizonless objects with a reflective surface: black stars \label{sec:blackstars}}

The idea that the so-called astrophysical black holes may correspond instead to some kind of ultra-compact star with no horizons has been put forward in several occasions (see, e.g., \cite{Visser2009,Visser2009b} and references therein). During many years, models elaborating on this possibility have been met with skepticism, in part due to the lack of observational consequences associated with them. However, these models will be of clear relevance in the era of gravitational wave astronomy \cite{Thorne1997}: the possible existence of echoes in the late-time ringdown of gravitational wave signals originated in mergers of astrophysical black holes constitute one of the first indications of their potential.

Another important reason behind their lack of popularity is precisely rooted in the fact that BUCHOs would necessarily convey non-perturbative deviations of the near-horizon external geometry black holes (as well as radical modifications of the internal geometry) which, as emphasized in Sec. \ref{sec:classification}, correspond to approaches in the type I and IIB categories. But this is precisely the property that makes them interesting from an observational perspective, as we are stressing in this paper.

BUCHOs have been typically associated with approaches falling into the type I category. We have explained in Sec. \ref{sec:classification} that there is arguably no solid physical and theoretical ground for type I effects to exist. However, this does not rule out the existence of BUCHOs, as we have also stressed that certain type IIB effects may lead to the formation of these structures. Hence in the following we consider the particular approach developed by the authors, in which the so-called black stars are the expected stable outcome \cite{Barcelo2010,Barcelo2014e,Barcelo2014,Barcelo2015u,Barcelo2015,Barcelo2016}.  

A black star is a BUCHO which is filled with matter and is supported by quantum gravity effects (possibly vacuum polarization). Following our description of type IIB category, these effects do not appear before the formation of the trapped surface (crossing of the gravitational radius) in the gravitational collapse of a compact star, but rather arise when Planckian densities are reached. However, their non-local nature lead to macroscopic modifications of the resulting geometry. This leads to a series of bounces in which trapped and anti-trapped surfaces are formed for short periods of time. Dissipation makes the maximum radius attained by the bouncing structure at each bounce to slowly approach its gravitational radius. After a number of bounces this would permit vacuum polarization effects to build up and eventually halt the collapse through the mechanism explained in \cite{Barcelo2007}, leading to the formation of a black star \cite{Visser2009}.
 
The first relevant observation for the present discussion is that, being filled with matter, it is natural to consider that the surface of the black star would have reflective properties. Indeed, this was already pointed out some years ago in \cite{Barcelo2010}, as well as the existence of echoes due to the reflection of incoming waves with timescales given by Eq. \eqref{eq:timescale}. 

It is far from trivial that a particular realization of BUCHOs has to present reflective properties. For instance, in the gravastar scenario proposed by Mazur and Mottola \cite{Mazur2004,Mottola2010}, one of the most popular type I scenarios \cite{Visser2003,Cattoen2005,Broderick2007,Rocha2008,Sakai2014}, the horizonless object is conceptualized as essentially vacuum, that is, not filled with matter. In stark difference with the black star scenario, the original proponents of this scenario consider that one of the observational opportunities to distinguish gravastars from black holes would be the fact that incoming waves travel through the structure, being deflected during their crossing time \cite{Mazur2015} (a possibility which is allowed by the hollow core of gravastars). This example makes particularly clear that different realizations of BUCHOs still have freedom to lead to different phenomenology and, in particular, that reflective properties in the relevant range of frequencies are a must in order to lead to gravitational echoes. It is worth stressing in this regard that gravastars may be also recognized by looking at the very form of the merger ringdown \cite{Chirenti2007,Chirenti2016} (and not only by checking the existence or not of echoes in the late-time ringdown).

On the other hand, the properties of matter in the highly dense configurations attained in a black star are still a matter of research, which means that the reflective properties of the surface for different ranges of frequencies are still to be determined. A recurrent criticism against black hole substitutes is that the existence of surfaces is constrained by astrophysical data that involve frequencies down to the infrared \cite{Broderick2009,Broderick2015}. While these observations greatly constrain objects such as boson stars outside the BUCHO category, it is not clear however how tight these constraints are (or can ever become) for BUCHOs \cite{Abramowicz2002} and, most importantly, whether these constraints suffice to discard the existence of all kind of surfaces without a thorough analysis of the physical properties of matter in these extreme situations. Pending a more detailed discussion of this issue (which, perhaps surprisingly, is lacking in the literature), one possible option might be that BUCHOs do not display reflective properties for any range of frequencies which, in particular, would discard as well the existence of gravitational echoes. 

A more interesting alternative is that the black star surface is reflective for low frequencies, while becoming perfectly absorbing for high frequencies. It would be interesting to find a universal argument that justifies these properties without the need to consider the fine details concerning the state of matter in these configurations (though perhaps this may be unavoidable at the end of the day). We find intriguing the following consideration in this regard, motivated by some comments in \cite{Abedi2016}. It has been extensively justified that black stars could emit Hawking-like radiation \cite{Barcelo2006b,Barcelo2010b,Barcelo2010c,Barbado2011}, something which would probably apply to other BUCHOs. This would associate an effective temperature to every black star which is essentially (up to small corrections) given by the Hawking temperature. Matter in black stars would represent the most dense stable states that are attainable in nature, so it is natural to associate them with some fundamental state of matter, even more if taking into account that macroscopic quantum effects play a crucial role in its stabilization. It is tempting to associate the Hawking temperature $T_{\rm H}$ to some measure of the width of a gap associated with the (de-)excitation of this state of matter. Then the surface of the star will behave as a filter that absorbs incoming wavelengths with frequency greater than a certain cutoff frequency $\nu_{\rm c}$, defined in terms of the frequency associated with the Hawking temperature ($k_{\rm B}$ is Boltzmann's constant)
\begin{equation}
2\pi\nu_{\rm H}=k_{\rm B}T_{\rm H}=\frac{1}{8\pi M}.\label{eq:cutoff}
\end{equation}
On the other hand, frequencies smaller than $\nu_{\rm c}$ will be reflected due to their energy being smaller than the spacing of energy levels. There is a large uncertainty in the particular value of the cutoff frequency $\nu_{\rm c}$, as it could be several orders or magnitude larger or smaller than $\nu_{\rm H}$, depending on the specifics of its definition (and also on the width of the transition region between the purely absorbing and reflective domains). This example just illustrates that it should be possible to obtain models of black stars that present gravitational echoes, that is, that have a reflective surface for gravitational waves of the expected frequencies in the events involved, while being able to absorb incoming waves with higher frequency, thus satisfying potential astrophysical constraints obtained through electromagnetic observations (let us stress again that a detailed study of the theoretical consequences of observational constraints using electromagnetic observations would be in place).

It might be instructive to complement this abstract discussion with some real numbers, taken for instance from the first gravitational wave observation of the merger of astrophysical black holes \cite{Abbott2016}. In this merger event, the resulting astrophysical black hole has a mass of $62M_{\odot}$. This leads to a frequency $\nu_{\rm H}=21$ Hz. On the other hand, frequencies registered in the signal go from 35 to 250 Hz. The relevant range of frequencies is close to the latter value (these are the frequencies produced in the ringdown phase), and hence will be reflected as long as the cutoff frequency is at least an order of magnitude greater than $\nu_{\rm H}$. While this example is too rough to take these numbers too seriously, it serves to stress however how crucially the existence of echoes relies on the (potentially frequency-dependent) reflective properties of BUCHOs, making clear the need for additional theoretical analyses of these reflective properties.

Our last comment regarding the black star scenario is about the transient leading to the final BUCHO state which, as stated in Sec. \ref{sec:classification}, is radically different in both type I and type IIB approaches. The very proposed dynamical mechanism for the formation of of black stars in type IIB approaches, namely the existence of a sequence of bounces before stabilization happens, leads itself naturally to the existence of reverberations (i.e., dampened repetitions of the primary signal) in gravitational wave signals \cite{Barcelo2014,Barcelo2015u}. The corresponding timescale in this case was shown to be slightly shorter than Eq. \eqref{eq:timescale} (it is linear in the Schwarzschild mass but with no multiplicative logarithms), and corresponds to a low-energy observable that should be calculable in the framework of quantum gravity \cite{Barcelo2016}. These reverberations will be manifest in different astrophysical events. Hence, irrespectively of observations of the stable end state, monitoring the formation of BUCHOs (if existent) will provide a precious information about the actual mechanism behind their formation.

\subsection{Black hole complementarity, firewalls, and fuzzballs \label{sec:firewalls}}

First of all, it is important for our discussion to stress that firewalls \cite{Almheiri2012,Almheiri2013,Braunstein2009} were originally formulated in the framework of black hole complementarity \cite{Susskind1993,Kiem1995}. One of the central hypothesis of this paradigm is that external observers can use a perturbative semiclassical description in order to describe exhaustively black hole physics in the external geometry that is accesible to them. Internal observers (observers falling into black holes and crossing the horizon) will be able to map the internal geometry all the way down close to the singularity, in which quantum gravity effects cannot be neglected. In black hole complementarity, the experiences of external and internal observers are not reconcilable, but this does not posit a logical problem as no observer will be able to get together the two versions of the story. In summary, this paradigm fits in the type IIA category introduced in Sec. \ref{sec:classification} (and is indeed one of the maximum exponents of it).  

Part of the complementarity paradigm is the existence of a membrane hovering over the horizon, but extremely close to it (by a distance of the order of Planck length). This membrane is assumed to be physical, but all its properties are designed in order to permit that external observers can describe all physical processes in the external black hole geometry using perturbative deviations from general relativity. In particular, this membrane should absorb any incoming form of energy. Departures from this behavior would contradict the overall motivation of the paradigm of black hole complementarity.

On the other hand, the complementarity paradigm asserts that observers crossing the horizon will not experience anything unusual and, in particular, will not detect any membrane. Equivalently, the description of the internal geometry is also perturbative from the classical description in general relativity (except close to the singularity in which curvatures are Planckian). This statement is the (only) one challenged by firewalls \cite{Almheiri2012,Almheiri2013}. The study of the behavior of entanglement across the horizon suggests that observers crossing the horizon should indeed see a wall of concentrated energy (or energetic curtain \cite{Braunstein2009}). This will happen only for black holes old enough, namely after the Page time $T_{\rm Page}\propto M^3$ (defined as the time in which a black hole halves its area due to the emission of Hawking evaporation). The ignition of firewalls is sometimes hypothesized to start around the scrambling time $T_{\rm Scr}\propto M\ln M$ \cite{Almheiri2012}, though there is no strong additional argument showing that this is indeed the case. Let us stress that the original argument pointing to the creation of firewalls for infalling observers due to the behavior of entanglement applies only after the Page time.

The firewall argument shows only that the ignition of the firewall for black holes older than the Page time has to affect drastically the internal geometry, even terminating spacetime (see \cite{Brustein2015,Brustein2015b} for some alternative pictures). There is no acknowledged reason that would make firewalls affect the external geometry and therefore the predictions made by external observers. Indeed, this would violate the principles of black hole complementarity (and would make firewalls no longer be the ``most conservative'' solution to the information loss problem, in the words of \cite{Chen2015}). It could be possible that firewalls become naked in certain intervals of time after their formation due to the evaporation of black holes, though this possibility is poorly understood \cite{Chen2015,Hwang2012,Kim2013}.

These considerations make firewalls fit in type IIA approaches, in particular breaking any potential link between firewalls and reverberations in gravitational wave signals from mergers. However, we can go further for the sake of completeness. Let us make the following assumption that represents already a huge leap from the original firewall picture sketched above, and accept that the ignition of firewalls can somehow affect the black hole membrane, making it temporarily reflective to incoming waves (why this should be the case, is a mistery). This assumption represents an unconventional distortion of the firewall concept, as there is no theoretical argument supporting that firewalls should affect the external geometry of a black hole (this distortion would now fit into type IIB category instead). What is clear from the firewall argument is that the formation of a firewall has to be an equilibrium property that arises some time after the black hole horizon (membrane) is settled. A simple analogy is that of a thin oil layer (the firewall) on a water surface (the black hole horizon/membrane). If the water surface is stirred violently (e.g., during and just after the merger of two black holes), the oil layer will be broken. If the water surface is left to come back to equilibrium, the oil layer will smoothly cover it again after some relaxation timescale.

The existence of this relaxation timescale connects naturally with our previous discussion about timescales at the beginning of Sec. \ref{sec:scales}, and in particular with the timescale of development of reflective properties $\mathscr{T}_{\rm s}$. As stated above, the finest arguments point to timescales of the order of the Page time. This time scale is huge when compared with Eq. \eqref{eq:timescale}, discarding again the existence of echoes even in this rather unconventional approach, as Eq. \eqref{eq:maincond} would be largely violated:
\begin{equation}
\mathscr{T}_{\rm s}\simeq T_{\rm Page}\gg 
\frac{\mathscr{T}_{\rm e}}{2}.
\end{equation}
Even if one considers that $\mathscr{T}_{\rm s}$ in this scenario lies between the Page time and the scrambling time, in general situations the inequality \eqref{eq:maincond} will not be satisfied. The evident problem is that the scrambling time is of the same order of magnitude than Eq. \eqref{eq:timescale}, which makes unnatural that reflective properties are completely developed fast enough. For any value $\mathscr{T}_{\rm s}\in[T_{\rm Scr},T_{\rm Page}]$ it would be
\begin{equation}
\mathscr{T}_{\rm s}\gtrsim
\frac{\mathscr{T}_{\rm e}}{2}.
\end{equation}
This equation has to be compared with Eq. \eqref{eq:maincond}.

The last paragraph in this section is devoted to the exotic objects known as fuzzballs \cite{Mathur2005}. For the sake of the present discussion it is enough to recall that fuzzballs are members of the type IIA family and, in particular, fuzzballs respect complementarity \cite{Mathur2011,Mathur2012,Mathur2012b,Mathur2013,Avery2012} (in more precise terms, the so-called ``fuzzball complementarity'', the difference of which with respect to the standard complementarity discussed above is not of relevance for the behavior of the external black hole geometry). Hence our general comments above concerning the lack of echoes in approaches in the type IIA category trivially applies to fuzzballs.

\section{Conclusion \label{sec:conc}}

Discussions aimed at establishing a coherent picture of black holes beyond general relativity have been historically driven by theoretical reasoning, and adherence to different school of thoughts were to a great extent a matter of taste (or scientific community of origin). It is remarkable that new observational windows, brought in particular by gravitational wave astronomy, may change this trend by making these choices no longer inconsequential from an observational perspective. In order to work towards this goal, in this paper we have cleared up the connection between fundamental assumptions lying at the core of different approaches and the phenomenological implications that may (or may not) be expected.

The aim of this paper is to set a classification scheme (type I, IIA and IIB approaches) that provides a frame of reference for discussions concerning the phenomenology of quantum gravity effects in black holes. The usefulness of this classification has been highlighted using a particular example, namely that of reverberations or echoes in gravitational wave signals from binary mergers of astrophysical black holes. In particular, we have used these categories to provide a clean argument that shows the absence of gravitational echoes in certain conceptualizations of black holes beyond general relativity.  We think this classification will remain useful in discussions about this and other examples of quantum gravity phenomenology that may be open for exploration in the (near) future.

\bibliography{refs}

\end{document}